\documentclass[fleqn,usenatbib]{mnras}

\usepackage{newtxtext,newtxmath}

\usepackage[T1]{fontenc}
\usepackage{ae,aecompl}

\usepackage{graphicx}	
\usepackage{amsmath}	
\usepackage{graphicx}	
\usepackage{amsmath}	
\usepackage[shortlabels]{enumitem}      
\usepackage{booktabs}   
\usepackage[nice]{nicefrac}  
\usepackage{mathtools}
\usepackage{bm}
\usepackage{esvect}
\usepackage{multicol}
\usepackage[export]{adjustbox}

\newcommand{\orange}[1]{{#1}}

\graphicspath{{Figures}}
\DeclareGraphicsExtensions{.png,.pdf,.eps}

\title[Halos and voids with stacked WL and DL]{Mapping the cosmic mass distribution with stacked weak gravitational lensing and Doppler lensing}

\author[Hossen et al.]{
Md Rasel Hossen$^{1}$\thanks{E-mail: mhos5289@uni.sydney.edu.au},
Sonia Akter Ema$^{1}$, 
Krzysztof Bolejko$^{2}$ and 
Geraint F. Lewis$^{1}$
\\
$^{1}$Sydney Institute for Astronomy, School of Physics, A28, The University of Sydney, NSW 2006, Australia \\
$^{2}$School of Natural Sciences, College of Sciences and Engineering, University of Tasmania, Private Bag 37, Hobart TAS 7001, Australia
}

\date{Accepted XXX. Received YYY; in original form ZZZ}

\pubyear{2021}

\begin{document}
\label{firstpage}
\pagerange{\pageref{firstpage}--\pageref{lastpage}}
\maketitle

\begin{abstract}
 Dark matter halos represent the highest density peaks in the matter distribution. Conversely, cosmic voids are under-dense patches of the universe. Probing the mass distribution of the universe requires various approaches, including weak gravitational lensing that subtly modifies the shape of distant sources, and Doppler lensing that changes the apparent size and magnitude of objects due to peculiar velocities. In this work, we adopt both gravitational and Doppler lensing effects to study the underlying matter distribution in and around cosmic voids/halos. We use the relativistic $N$-body code \texttt{gevolution}, to generate the mass perturbations and develop a new ray-tracing code that relies on the design of the ray bundle method. We consider three categories of halo masses and void radii, and extract the cosmological information by stacking weak-lensing and Doppler lensing signals around voids/halos. The results of this paper show that the most optimal strategy that combines both gravitational and Doppler lensing effects to map the mass distribution should focus on the redshift range $z\approx 0.3-0.4$. The recommendation of this paper is that future spectroscopic surveys should focus on these redshifts and utilise the gravitational and Doppler lensing techniques to extract information about underlying matter distribution across the cosmic web, especially inside cosmic voids. This could provide a complimentary cosmological analysis for ongoing or future low-redshift spectroscopic surveys.
\end{abstract}

\begin{keywords}
gravitational lensing: weak-- cosmology: large-scale structure of universe-- cosmology: dark matter-- methods: numerical
\end{keywords}



\section{INTRODUCTION}
\maketitle

Our current universe appears to be quite clumpy, with countless galaxies, groups of galaxies, galaxy clusters, and so on, spanning a wide dynamic range in mass. The mass distribution of the large-scale structure (LSS) is a key factor in understanding the evolution of the universe. Due to the variations of the mass distributions of matter in the LSS: galaxies and galaxy clusters correspond to over-dense regions \citep{Broadhurst2005, Umetsu2011} while cosmic voids are the under-dense regions \citep{Park2007}. Over the last few decades a substantial amount of research works have been published with cosmic voids by utilizing the effects of void ellipticities, void size functions, the emptiness and its evolution over cosmic time, etc. \citep{Park2007, Bos2012, Pan2012, Krause2012, Sutter2012, Bolejko2013, Ceccarelli2013, Hamaus2014, Nadathur2014, Hamaus2015, Nadathur2016, Sanchez2016, Mao2017, Verza2019, Panchal2020, Li2020, Raghunathan2020} and halos by analysing the mass distributions of galaxies and galaxy clusters, sizes, and magnitudes of galaxies, etc. \citep{Kaiser2000, Schmidt2012, Han2015, Fong2018, Salcedo2019} for constraining cosmology and extracting the information from the expansion history of the universe and modified gravity \citep{Farrar2004, Peebles2010}.

The weak gravitational lensing technique is a very common method to investigate the matter distribution. Weak gravitational lensing, the distortion of images when light bundles are deflected by the mass distribution of LSS as they travel the tremendous distance from the source to the observer, provides the most efficient way of measuring the distribution of matter density in the universe \citep{Kaiser1992, Kaiser1993, Mellier1999, Bacon2000, Wittman2000, Bartelmann2001}. The distortion patterns in the images from distant galaxies provide a powerful window for studying the expansion of the universe and the evolution of the cosmic structure \citep{Wambsganss1998, Wittman2000, Kilbinger2015}. Ongoing surveys such as the Dark Energy Survey (DES\footnote{\orange{http://www.darkenergysurvey.org}}), the Dark Energy Spectroscopic Instrument (DESI\footnote{\orange{https://www.desi.lbl.gov/}}) Bright Galaxy Survey \citep{Ruiz-Macias2020, Zarrouk2021}, and Hyper-Suprime-Cam (HSC\footnote{\orange{http://hsc.mtk.nao.ac.jp/ssp/}}) or future weak-lensing surveys such as Euclid\footnote{\orange{https://sci.esa.int/web/euclid}}  and the Large Synoptic Survey Telescope \citep{LSST2012} aim to cover a wide portion of the sky with great precision.
The weak-lensing signal for single structures can be very difficult to measure. One of the reasons is the noise due to foreground structures also contributing to lensing distortions. This can be minimised by the method of stacking weak-lensing signal produced by the large number of targeted structures. This can reduce the errors of the statistical analysis, which introduce a powerful tool to probe the average mass distribution of the LSS for understanding the evolution of the universe \citep{Oguri2011, Higuchi2013, Simet2015}.

The mapping of the distribution of galaxies and clusters of galaxies (by neglecting all peculiar velocities) observed in redshift-space would be identical as compared with the real space in a perfectly homogeneous universe \citep{Kaiser1987}. But the peculiar velocities of galaxies associated with any inhomogeneous structure will introduce distortion in this mapping along the line of sight \citep{Davis1983,Kaiser1987}. This distortion effect is known as redshift-space distortion. This effect has been studied extensively in both linear \citep{Davis1983,Kaiser1987} and nonlinear \citep{Cole1995, Magira2000} regimes.

A complimentary method to gravitational lensing is the Doppler lensing. The Doppler lensing signal is caused by the redshift distortions. These distortions are then expressed in terms of weak-lensing quantities \citep{Bolejko2013, Bacon2014, Bonvin2017}. Since the redshift distortions are caused by the underlying matter distribution, the Doppler lensing offers a complimentary approach to investigate matter distribution. At low redshift, the Doppler lensing is much stronger than weak-lensing,  and thus convergence is dominated by the Doppler lensing \citep{Bolejko2013, Bacon2014}. Thus the correct method of analysing lensing statistics needs to incorporate both: gravitational and Doppler lensing.

\citet{Bacon2014} suggested a new approach for detecting Doppler magnification that incorporated cross-correlating the convergence by using galaxy sizes and magnitudes. They used a Newtonian simulation and stacked the Doppler lensing signal around the cosmic voids and halos. They showed that at low redshift Doppler lensing dominates over standard weak-lensing and at higher redshift Doppler lensing falls while weak-lensing grows.  They also constructed an estimator based on angular power spectrum which can be used to constrain the cosmological model. Later on, \citet{Bonvin2017} investigated the dipolar modulation in the size of galaxies by using the effect of Doppler magnification. They showed that by extracting dipole in the cross-correlation of number counts and galaxy sizes one can able to detect the Doppler lensing in current and upcoming radio surveys.

The formation of the LSS is complex because it becomes non-linear as the inhomogeneities grow, so numerical methods are required. Over the last few decades, significant attention has been devoted to study the weak gravitational lensing by the LSS associated with numerical simulations, mostly based on Newtonian gravity \citep{Jain2000, Barber2003, Hilbert2009, Takahashi2011, Killedar2012, Valageas2012}, but where the presence of relativistic sources is prevalent, this approach has some limitations to study the formation of cosmic structures \citep{Chisari2011, Green2012, Adamek2013, Lepori2020}. As the cosmic shear signal provides the most useful cosmological information in the small angular scales so, not only the nonlinear effect but also including the relativistic effect (by implementing the general relativity treatment for gravity instead of Newtonian gravity) provides more accurate modelling of the weak gravitational lensing statistics. Recently, \citet{Adamek2016a, Adamek2016b} developed a new $N$-body code, namely \texttt{gevolution}\footnote{\orange{https://github.com/gevolution-code/gevolution-1.2}}, based on the theory of general relativity, and our present work will use this $N$-body simulation to study the relativistic weak-lensing statistics. In this paper we use \texttt{gevolution} and adopt the ray-tracing framework developed in our companion paper \citep{Ema2021} (which focuses on studying the impact of the local environment on weak-lensing statistics).

In this paper, to study the effect of the underlying matter distribution in and around cosmic voids/halos, we will adopt both Doppler lensing and standard weak gravitational lensing and analyse their properties by utilizing the information from the statistical quantities i.e. convergence, shear, and magnification. Here we neglect the local environment effect on light propagation. We will show how the properties of Doppler lensing differ from those of standard weak-lensing and how the distribution of matter in the universe can impact the weak-lensing signal.

The layout of this paper is as follows: in Section \ref{background}, we present the background of weak gravitational lensing, Doppler lensing and numerical modelling of our ray tracing algorithm, and description of the $N$-body simulations. We show our results and analysis of weak-lensing signal for clustering masses of the halos and clustering radii of the voids, and also show the comparison between the standard weak gravitational lensing and Doppler lensing in Section \ref{res}. 
Finally, we conclude in Section \ref{conclusion}.

\section{Background and Modelling}\label{background}

\subsection{Lensing}

\subsubsection{Weak gravitational lensing}
Here our aim is to briefly describe the theory of weak gravitational lensing which arises from the distortion of background galaxy shapes due to the intervening LSS from source galaxy to observer. It can be described by the Jacobian matrix mapping source angular positions $\boldsymbol{\theta}^\mathrm{S}$ to image positions $\boldsymbol{\theta}^\mathrm{I}$, i.e., $d\theta_i^\mathrm{S}=\mathcal{A}_{ij}d\theta_j^\mathrm{I}$. In the weak-lensing limit, this distortion matrix \citep{Bartelmann2001} can be decomposed as 
\begin{align}
\mathcal A=\left( \begin{array}{c c}
1-\kappa-\gamma_1 & -\gamma_2 \\
-\gamma_2 & 1-\kappa+\gamma_1\\
\end{array} \right),
\end{align}
which defines the convergence field $\kappa$ and complex shear field $\gamma\equiv \gamma_1+i\gamma_2$. The total magnification of surface area elements $\mu$ is given by the determinant of the inverse matrix
\begin{equation}
\mu = \frac{1}{\det(\mathcal A)}=[(1-\kappa)^2-|\gamma|^2]^{-1},\label{mag}
\end{equation}
which in the weak-lensing limit $|\kappa|,\,|\gamma|\ll 1$ can be approximated by $\mu\simeq 1+2\kappa$. The convergence field $\kappa$ tells us at what amount a source galaxy at a fixed redshift is magnified or demagnified by the intervening of the LSS while the shear field $\gamma$ is the stretching along the axes of the image.  The tangential shear changes the orientation of the background source while convergence is responsible for the magnification or de-magnification of the source. For an azimuthally symmetric lens the convergence $\kappa$ and the tangential shear $\gamma$ and  can be expressed as
\begin{equation}\label{eq:proj} 
\kappa = \frac{\Sigma(r)}{\Sigma_{\rm cr}} \quad {\rm and} \quad \gamma =\: \frac{\Delta\Sigma(r)}{\Sigma_{\rm cr}},
\end{equation}
\noindent where $\Sigma(r)$ represents the projected surface mass density of the lens, $\Delta\Sigma(r)$ is the differential surface mass density of the lens, $r$ is the physical transverse distance on the lens plane
\begin{equation}
\Delta\Sigma(r) =\: \bar\Sigma(< r) - \Sigma(r),
\end{equation}
\noindent  and $\Sigma_{\rm cr}$ is defined as 
\begin{equation}\label{eq:Scr}
\Sigma_{\rm cr} =\: \frac{c^2}{4\pi G}\frac{D_s}{D_L\,D_{Ls}}, 
\end{equation}
\noindent where $D_s$ is the angular diameter distance to the source, $D_L$ is the angular diameter distance to the lens and  $D_{Ls}$ denotes the angular diameter distance between lens and source. Note that the differential surface mass density $\Delta\Sigma(r)$ is proportional to the weak-lensing tangential shear, while the projected surface mass density $\Sigma(r)$ is proportional to the weak-lensing magnification.

\subsubsection{Doppler lensing}

The relevance of peculiar velocities on apparent size and magnitude of observed objects was first highlighted by \cite{Bonvin2008}.  Subsequently, \cite{Bolejko2013} demonstrated that the effect of the Doppler lensing can dominate at low redshifts. The convergence due to Doppler lensing can be expressed by the following normalised equation
\begin{equation}
\kappa_v=\left(1-\frac{1+z_s}{ H\chi_s}\right) \bm{v}/c\cdot\bm n,
\label{eqn:Dop}
\end{equation}
\noindent where $z_s$ is the source redshift, $H$ is the Hubble parameter, $\chi_s$ is the co-moving distance of the sources, \bm{$v$} is the velocity of the source galaxies, $c$ is the speed of light, and \bm{$n$} is the unit vector from the source to the observer (the directions in which photon propagates). Here, the sign of convergence due to Doppler lensing, $\kappa_v$ alters according to whether objects travel towards or away from us. More specifically, the term $\bm v/c\cdot\bm n > 0$ ($\bm v/c\cdot\bm n < 0$) indicates that the object travel towards (away) from us. The `+ve' (`-ve') sign of the value of Doppler convergence implies that at their observed redshift they are smaller and dimmer (larger and brighter) than typical objects. While weak gravitational lensing is caused by the mass distribution, Doppler lensing is an effect that is purely related to the peculiar velocity of objects. We will discuss more the effect of the Doppler lensing, and compare with standard weak gravitational lensing in Section \ref{res:Dop}.

\begin{figure*}
    \centering
    \hspace*{-0.3in}
    \includegraphics[width=2.2\columnwidth]{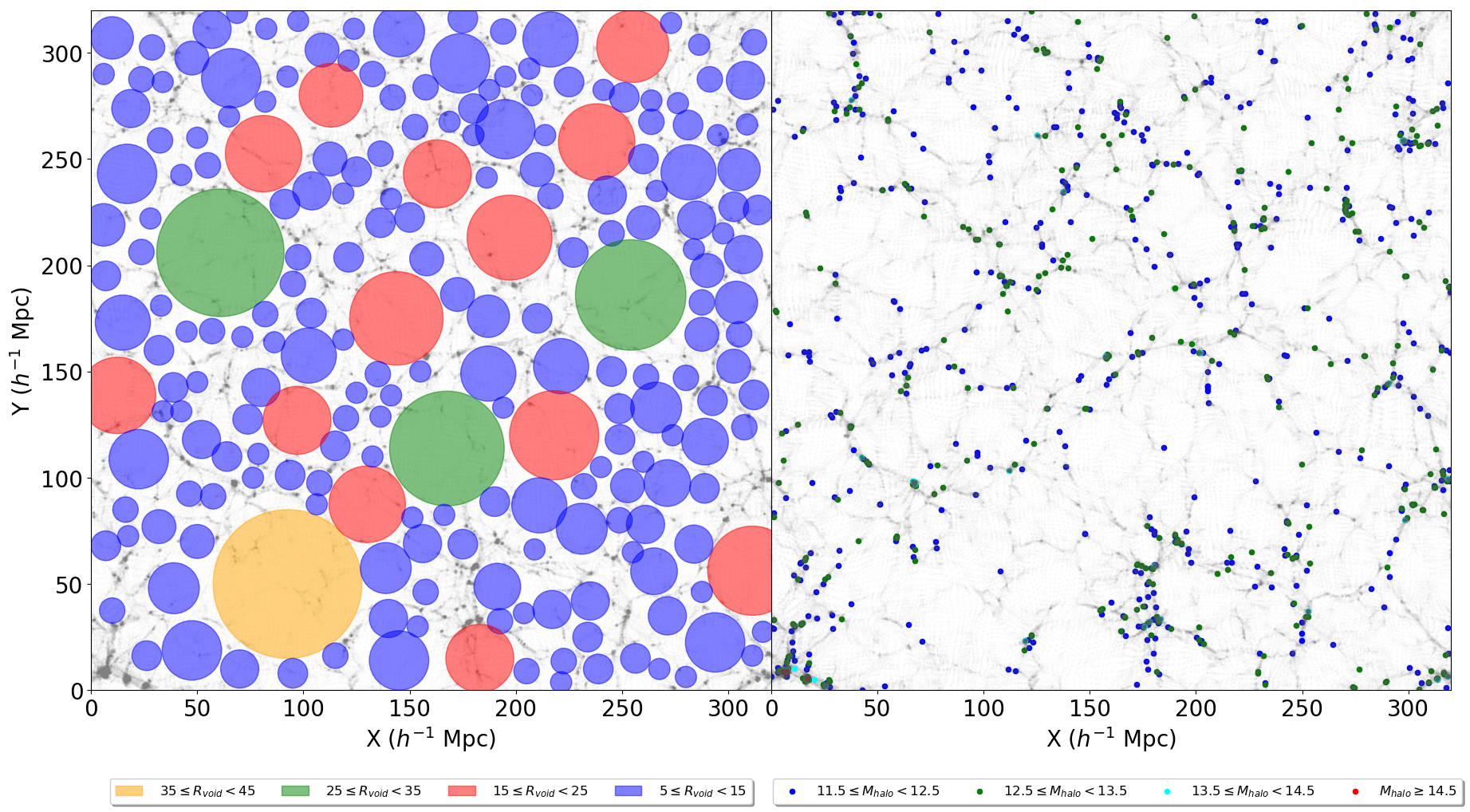} 
    \vspace*{-6mm} 
    \caption{The structure of the cosmic voids and halos in a relativistic cosmological simulation. The left panel of the figure shows the clustered voids having different ranges of radii within a $10~h^{-1}\mathrm{Mpc}$ slice of the \texttt{gevolution} simulation at $z=0.22$. The grey colour indicates the DM distributions whereas the different colour circles are the voids of different ranges of radii in units of $h^{-1}\mathrm{Mpc}$. The right panel of the figure shows the clustered halos having different ranges of masses (all values in the right side legends indicate in the order of 10 magnitudes and in units of $~h^{-1}M_\odot$) within a $10~h^{-1}\mathrm{Mpc}$ slice of the \texttt{gevolution} simulation at $z=0.22$. The grey colour indicates the DM distributions whereas the different colour dots are the halos of different ranges of masses. }
    \label{fig2}
\end{figure*}

\begin{figure}
    \hspace{-0.3cm}
    \centering
    \includegraphics[width=\columnwidth]{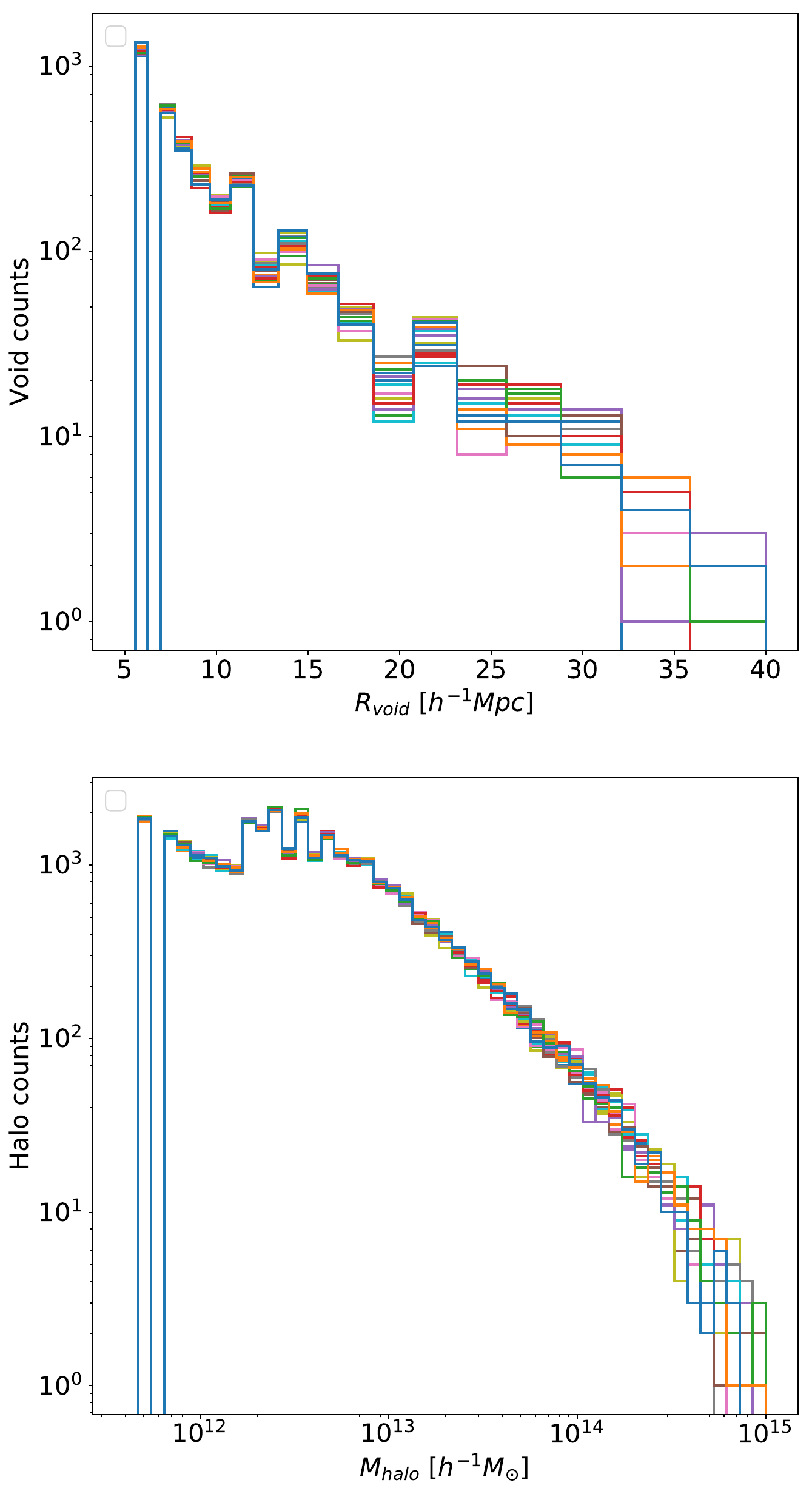}
    \caption{Variations of the number of halo and void counts for 21 realisations of \texttt{gevolution} simulations at redshift $z=0.22$. The top panel shows the number of void counts as a function of the radii of voids and the bottom panel shows the number of halo counts as a function of the masses of halos. }
    \label{new_fig}
\end{figure}

\subsection{Numerical simulations}

The aim of this paper is to investigate the applicability of the gravitational and Doppler lensing methods.  The goal is to obtain mass profiles of cosmic structure. We test the method using relativistic cosmological simulations from \texttt{gevolution} and identify cosmic structures from the publicly available codes \texttt{ROCKSTAR} \citep{Behroozi2013} and \texttt{Pylians} \citep{Villaescusa-Navarro2018}. There are some other robust void finding codes \citep{Nadathur2015a, Nadathur2015b, Sutter2015} based on the parameter-free void finding algorithm \texttt{ZOBOV} \citep{Neyrinck2008} but in this work, we will use \texttt{Pylians} to find the spherical underdense regions.

\subsubsection{Metric}

Here our goal is to discuss relativistic $N$-body simulations, and some existing methods for the ray-tracing algorithm to analyse the weak-lensing statistics. 
Both, simulations and light propagation methods are based on the perturbations around the homogeneous isotropic FLRW universe, with the metric in the conformal Poisson gauge give as 

\begin{align}\label{e:metric}
ds^2 = a^2(\tau) [-(1+2\Psi)d\tau^2 - 2B_ix^id\tau + (1-2\Phi)\delta_{ij}dx^idx^j \notag\\
+ h_{ij}dx^idx^j],
\end{align}
\noindent where $a$ denotes the scale factor of the background, $x^i$ are the comoving Cartesian coordinates, and $\tau$ is the conformal time. $\Phi$ and $\Psi$ are the scalar perturbations, $B_i$ is the vector perturbation and $h_{ij}$ is the tensor perturbation which contain two spin-2 degrees of freedom of gravitational waves. In  \texttt{gevolution} simulation, the coordinate system is fixed by the gauge conditions
\begin{equation}
 \delta^{ij} B_{i,j} = \delta^{ij} h_{ij} = \delta^{jk} h_{ij,k} = 0 .
\end{equation}

The two scalar perturbations ($\Phi$ and $\Psi$) from \texttt{gevolution} having the value of the order of $10^{-5}$ which is at least two orders of magnitude higher than the vector potentials and the tensor perturbation $h_{ij}$ is even smaller on the scales that we are interested in \citep{Lu2009, Thomas2015, Adamek2016a, Lepori2020}. In our first companion paper \cite[see Fig. 2 of][]{Ema2021}, we illustrate that the two scalar perturbations ($\Phi$ and $\Psi$) contribute the maximum effect on the light propagation and the vector potential has negligible effect on weak-lensing statistics, hence we neglect the vector and tensor perturbations contribution throughout this work.

\subsubsection{\texorpdfstring{$N-$}-body simulation}
To study the weak-lensing statistics, the weak potentials are generated from a relativistic $N$-body simulations and a set of null geodesic equations are integrated through a cosmological mass distribution. Our results are based on relativistic $N$-body simulation has $256^3$ mass particles for a cosmological volume of $(320~\mathrm{Mpc}/h)^3$ carried out with the relativistic code \texttt{gevolution}. The metric is sampled on a regular Cartesian grid of $256^3$ points, providing a spatial resolution of $1.25~h^{-1}\mathrm{Mpc}$. This allows robust detection of dark matter (DM) halos down to about $5 \times 10^{11}~M_\odot/h$ (we find it by using \texttt{ROCKSTAR}). We choose a baseline $\Lambda$CDM cosmology with $h = 0.67556$, $\Omega_c = 0.2638$, $\Omega_b = 0.048275$, and a radiation density that includes massless neutrinos with $N_\mathrm{eff} = 3.046$. Linear initial conditions are computed with \textit{CLASS} \citep{Blas2011} at redshift $z_\mathrm{ini} = 127$, assuming a primordial power spectrum with amplitude $A_s = 2.215 \times 10^{-9}$ (at the pivot scale $0.05~\mathrm{Mpc}^{-1}$) and spectral index $n_s = 0.9619$. 

In this paper we consider a volume of approximately $\left( 900~h^{-1} {\rm Mpc} \right)^3$ (volume equivalent to a spherical slice between $z=0.2$ and $z= 0.25$). 
This volume was simulated with the means of 21 realisations of $320~\mathrm{Mpc}/h$. 
The computational time for 21 realisations of \texttt{gevolution} simulations on the above configurations takes about 250 CPU hours. 
The reason for the simulation of the required volume with 21 realisations instead of a single larger box was motivated by the requirements of the ray-tracing algorithm. The ray-tracing algorithm is the bottleneck of our pipeline, which requires substantive quantities of CPU time and large memory. The trade-off of this approach is the lack of extremely large structures and the need of having rays crossing the boundaries between boxes.

\subsubsection{Halo finding}
To identify the DM halos, we use the publicly available code \texttt{ROCKSTAR}\footnote{\orange{https://bitbucket.org/gfcstanford/rockstar/src/main/}}, a phase-space friends-of-friends (FOF) halo finder. We run \texttt{gevolution} having a cosmological volume of $(320~\mathrm{Mpc}/h)^3$ and read the particle snapshots at $z=0.22$ by using \texttt{ROCKSTAR} which gives more than $3 \times 10^4$ DM halos with a minimum number of particles considered to be a halo seed is 20 per halo \cite[by setting FOF refinement fraction = 0.7 and 3D-FOF linking length = 0.28, for more details about these please see][]{Behroozi2013}. This sets our minimal halo mass (halo radius) to $M_{halo, min}= 5 \times 10^{11}~h^{-1}M_\odot$ ~ ($R_{halo, min}=0.15~h^{-1}\mathrm{Mpc}$) and the maximal halo mass (halo radius) to $M_{halo, max}= 1.6 \times 10^{15}~h^{-1}M_\odot$ ~ ($R_{halo, max}=2.56~h^{-1}\mathrm{Mpc}$). To analyse the effect of the mass of halos ($M_{{\rm halo}}$), we subdivide the halos into four different mass ranges: i) $10^{11.5}~h^{-1}M_\odot < M_{halo} < 10^{12.5}~h^{-1}M_\odot$ ii) $10^{12.5}~h^{-1}M_\odot < M_{halo} < 10^{13.5}~h^{-1}M_\odot$ iii) $10^{13.5}~h^{-1}M_\odot < M_{halo} < 10^{14.5}~h^{-1}M_\odot$, and iv) $M_{halo} > 10^{14.5}~h^{-1}M_\odot$. Figure \ref{fig2} (right panel) shows the different clustering ranges of halo masses and the particle density distribution within a $10~h^{-1}\mathrm{Mpc}$ slice of the \texttt{gevolution} simulation at $z=0.22$. The redshift of $z = 0.22$ was chosen on the basis of the relevance of the Doppler lensing, which dominates for $z<0.3$ \citep[][cf. Sec. \ref{conclusion} for a further discussion]{Bacon2014}. The different colour dots indicate the halos having different masses and the legends represent the value of the ranges of the halo masses in units of $~h^{-1}M_\odot$. We find that the massive halos are located in the most clustered zone of the density of the particle distribution whereas the less massive halos are located in the less clustered zone of the density of the particle distribution. The fluctuation of the number of halo counts as a function of the masses of halos for 21 realisations of \texttt{gevolution} simulations at redshift $z=0.22$ is shown in the bottom panel of Fig. \ref{new_fig}.

\subsubsection{Void finding}
We employ \texttt{Pylians} \citep{Villaescusa-Navarro2018}, a publicly available python code, to find the cosmic voids. It starts with calculating a smoothly-varying density field from the particle positions within the gadget output of the \texttt{gevolution} simulation. We run \texttt{gevolution} having a cosmological volume of $(320~\mathrm{Mpc}/h)^3$ and read the particle snapshots at $z=0.22$ by using \texttt{Pylians}\footnote{\orange{https://github.com/franciscovillaescusa/Pylians3}} which gives more than $5.5 \times 10^3$ cosmic voids (by setting the threshold to  $-0.5$, which identifies voids with mean overall density contrast below this threshold). The code \texttt{Pylians} assumes the cosmic void as a sphere and the centre of this sphere is used as the void centre. It provides the information of the centres of the void positions and the radii of the cosmic voids. This sets the minimal cosmic void radius to $R_{void, min} = 5~h^{-1}\mathrm{Mpc}$ and the maximal void radius to $R_{void, max} = 44~h^{-1}\mathrm{Mpc}$. Similar to halos, we also divide four regions of voids having different ranges of radii: i)  $5~h^{-1}\mathrm{Mpc} \leq R_{void} < 15~h^{-1}\mathrm{Mpc}$ ii)  $15~h^{-1}\mathrm{Mpc} \leq R_{void} < 25~h^{-1}\mathrm{Mpc}$ iii)  $25~h^{-1}\mathrm{Mpc} \leq R_{void} < 35~h^{-1}\mathrm{Mpc}$, and iv) $35~h^{-1}\mathrm{Mpc} \leq R_{void} < 45~h^{-1}\mathrm{Mpc}$. The clustering range of voids having different radii and the particle density distribution within a $10~h^{-1}\mathrm{Mpc}$ slice of the \texttt{gevolution} simulation at $z=0.22$ is depicted in Fig. \ref{fig2} (left panel). The different circles with different colours indicate the voids having different ranges of radii and the colour bar shows the value of the ranges of the void radii in units of $~h^{-1}\mathrm{Mpc}$. As expected the voids are situated where the particle density distribution is low in our cosmological simulation volume. We demonstrate the variation of the number of void counts as a function of the radii of voids for 21 realisations of \texttt{gevolution} simulations at redshift $z=0.22$ in the top panel of Fig. \ref{new_fig}.

\subsection{Ray-tracing algorithm} \label{ray_alg}

The ray-tracing algorithm implemented in this paper is based on solving null geodesics equations. The metric  coefficients are obtained directly from relativistic simulations.
From these we calculate the Christoffel symbols and explicitly solve the null geodesics equations using the python package \textit{scipy.odeint}\footnote{\orange{https://docs.scipy.org/doc/scipy/reference/generated/scipy.integrate.odeint.html}} which uses the classical Runge-Kutta (RK-45) integration scheme. To fulfill the condition for light-like geodesic we check the null condition at each step of the integration. In order to minimise deviations from the null condition, we choose small values of $10^{-10}$ and $10^{-12}$ for the relative and absolute tolerance of the integrator, respectively.

\subsubsection{Weak gravitational lensing}\label{ray_alg1}

The statistical behaviors of gravitational lensing have been extensively studied by many researchers \citep{Schneider1988, Paczynski1989, Futamase1989, Lee1990, Holz1998, Killedar2012}. \cite{Watanabe1990} solved the relativistic optical equations to obtain the realistic distance-redshift relation in an inhomogeneous universe and observed that the influence of shear along the line-of-sight is modest when the scale of inhomogeneities is larger than the galactic scale. Recently, \citet{Adamek2019} also solved the relativistic optical equations and analysed the bias of the Hubble diagram from cosmological LSS. By using a Newtonian N-body simulation, \citet{Tomita1998} numerically integrate the null geodesic equations and observed that the different cosmologies have different angular diameter distances. \citet{Killedar2012} developed a three-dimensional ray-tracing algorithm by solving null geodesic equations and analysed the probability distribution function (PDF) of weak-lensing magnification and shear due to the mass distribution of source redshift of $z_s = 0.5$. By using a statistical method, \citet{Kaiser1992} and \citet{Jain1997} proposed a new measurement of the two-point correlation function and power spectrum due to the weak gravitational lensing of distant galaxies. To investigate the weak-lensing convergence and shear, \citet{Jain2000} developed a ray-tracing algorithm by considering ray shooting methods (RSM). Their algorithm follows the multiple-lens-plane method and found that the PDF and power spectrum of convergence are sensitive to the matter density. \citet{Hamana2000} also considered the multiple-lens-plane method and analysed the effect of weak-lensing magnification bias on the luminosity function of high-redshift quasars. The following is a list of the numerous methodologies that have been suggested for assessing cosmic properties: 
\begin{itemize}
    \item optical scalar methods: by solving optical scalar equations \citep{Kantowski1969, Dyer1974, Watanabe1990, Nakamura1997, Hamana1999, Adamek2019}.
    
    \item null geodesic methods: by integrating null geodesic equations backward from the observer to the sources where the weak perturbations are generated from $N$-body simulations \citep{Tomita1998, Killedar2012, Lepori2020}.
    
    \item statistical methods: measurements of the correlation function for ellipticity, and the corresponding angular power spectrum \citep{Blandford1991, Kaiser1992, Jain1997, Metcalf1999}.
    
    \item multiple-lens-plane methods: by considering a finite number of planes normal to light rays between an observer and a source, most commonly ray shooting methods \citep[RSM][]{Schneider1988, Paczynski1989, Lee1990, Jaroszynski1990, Hamana2000, Jain2000}. 
\end{itemize}

There are different methods that are available to investigate weak-lensing statistics. In this paper we adopted the ray bundle method  \citep[RBM][]{Fluke1999, Fluke2002, Fluke2011}. This method is similar to the RSM, but the methodology is preferred by other backward-ray-tracing codes \citep{Jain2000, Premadi2001} instead of using grid-based techniques to measure weak-lensing statistics around the source plane. For a pedagogical comparison between RBM and RSM for weak-lensing magnification, we refer the reader to Fig. 6 of \citet{Fluke1999}. 
The RBM considers bundle of light rays instead of a single light ray and each bundle consists of eight light rays with a central light ray \citep{Fluke1999, Fluke2002}. The advantage of using a bundle of light rays rather than a single light ray is that our algorithm can calculate both magnification and shear signals while other methods use a different technique to compute the shear signal. From the deformation of the shape of the bundle, we can calculate the weak-lensing statistical quantities i.e. convergence, shear, and magnification. In this work, a ray-tracing algorithm is developed that relies on the design of the RBM, and the photon path can be obtained by integrating a set of null geodesic equations backward from observer to source.

In our ray-tracing algorithm, instead of projecting a single photon (as RSM does), we have projected a bundle of photons (consisting of a central null geodesic surrounded by eight null geodesics) having a circular shape. As photon travels larger distances throughout the mass distribution of LSS in the universe, the circular shape of the bundle will be distorted because of the magnification (magnified or de-magnified) and shear (stretching along an axis). As we are integrating backward from observer to source the magnification ($\mu$) and shear($\gamma$) can be calculated as
\begin{eqnarray}\label{mu_eq}
\mu = \frac{A_i}{A_s} \quad \& \quad \gamma = \frac{b-a}{b+a},
\end{eqnarray}
\noindent where $A_s$ represents the area of the source, $A_i$ is the area of the image, a \& b are the semi-major and semi-minor axes of an ellipse, respectively. From the magnification and shear information, we can calculate the weak-lensing convergence ($\kappa$) by re-writing Eqn. \ref{mag} as

\begin{eqnarray}\label{kappa_eq}
\kappa = 1 - \sqrt{\frac{1}{\mu} + \gamma^2}.
\end{eqnarray}

To avoid considering a large simulation box (which is computationally time-consuming), we take the comoving length of our simulation box to be $320~h^{-1}\mathrm{Mpc}$. At first we save the scalar potentials and particle snapshots for different redshift by running the \texttt{gevolution} simulation (ensuring that the comoving length of each box to be $320~h^{-1}\mathrm{Mpc}$). When a photon reaches the edge of the box then we read another snapshot at the corresponding redshift having the same comoving length and take the trilinear interpolation scheme to calculate the value of the potential (as we use periodic boundary conditions for the \texttt{gevolution} simulation) so that at each step of the integration it will update the Christoffel symbols. Note that as the photon passes from box to box, the final direction of the exiting photon is the same as the initial direction of the entering photon, ensuring that the path of photon between the observer and the source is consistent.  Note that the ray-tracing algorithm we developed can project the bundle of light rays from any position of our simulation box. For more details about our ray-tracing algorithm we refer the reader to our companion paper \citep{Ema2021}.

To compute the weak-lensing statistics by stacking cosmic voids and halos, we follow the following steps: 

\begin{enumerate}[i]
\item at first we run 21 gevolution simulations (having identical cosmological parameter settings, but use a different random seeds) by changing the initial conditions to get the particle snapshots, and then by using \texttt{ROCKSTAR} and \texttt{Pylians} we find out the necessary information of the halos and voids at redshift $z= 0.22$; 
\item the choice of 21 of snapshots was motivated by the need to simulate approximately the volume of $\left( 900~h^{-1} {\rm Mpc} \right)^3$, i.e. equivalent of the volume of a slice between $z=0.2$ and $z= 0.25$ (as this is more relevant for the Doppler lensing than for gravitational lensing we comment on it in the next subsection);
\item we keep the observer fix at $z=0$, and by slightly varying the projection angles we project bundles of photons in the directions of pre-selected cosmic structures;
\item 
the structures are selected as follows: for voids we consider:
three different ranges of radii: $5~h^{-1}\mathrm{Mpc} \leq R_{{\rm void}} < 15~h^{-1}\mathrm{Mpc}$, $15~h^{-1}\mathrm{Mpc} \leq R_{{\rm void}} < 25~h^{-1}\mathrm{Mpc}$ and $25~h^{-1}\mathrm{Mpc} \leq R_{{\rm void}} < 35~h^{-1}\mathrm{Mpc}$), and for halos three different ranges of masses: $ M_{{\rm halo}} < 10^{12.5}~h^{-1}M_\odot$,  $10^{12.5}~h^{-1}M_\odot \leq M_{{\rm halo}} < 10^{13.5}~h^{-1}M_\odot$ and $M_{{\rm halo}} \geq 10^{13.5}~h^{-1}M_\odot$);
\item for each category of void radii and halo masses, we only select 100 objects (i.e. voids/halos). For each object void/halo, we solve 10000 bundles of geodesics, which is sufficient to obtain a clear signal and minimises time and memory issues with a larger number of object;
 \item when the light bundle exits the simulation box it enters it on the other side. Due to small angular scales of observed structures (i.e. we investigate lensing of particular structures rather than the cosmological signal of lensing power spectrum), low redshift, and the fact that bundles are shot at an angle to the side of the box, the bundles do not encounter the same structures on their way to the observed voids/halos (i.e. it only takes 4 boxes to reach $z=0.44$ and 6 to reach $z = 0.6)$; 
\item we keep our observer fix at the centre of our simulation box and determine the normal vectors from the observer's location to each projection direction using the Euler-Rodrigues formula. Each of these normal vectors corresponds to the initial direction of a particular ray bundle;
\item from the solution of the geodesics we get the positions of the bundles of photons for each integration step and to fit it into an ellipse we use an ellipse fitting algorithm to compute the area, semi-major and semi-minor axes of an ellipse;
\item then we compute the magnification from the ratio of the area of the final shape of the bundle to the area of the initial shape of the bundle since we integrate backward from the observer to the sources; 
\item from the information of the semi-major and semi-minor axes of an ellipse, we compute the shear by using Eqn. \ref{mu_eq} and then we compute the weak-lensing convergence by using Eqn. \ref{kappa_eq};
\item finally, we stack weak-lensing magnification and shear for voids having different ranges of radii (100 voids for each radius range) and halos having different ranges of masses (100 halos for each mass range), the results are presented in Figs. \ref{fig3} and \ref{fig4}.
\end{enumerate}

\begin{table*}
\begin{center}
\caption{Details of the number of halos and voids from 21 realisations of \texttt{gevolution} simulations at redshift $z=0.22$. The information of voids and halos are extracted from the publicly available codes \texttt{Pylians} and \texttt{ROCKSTAR}, respectively.}
\label{tab:all_void_halo}
\begin{tabular}{cccccc}
\hline
Halo mass [$h^{-1}M_{\odot}$] & No. of halos & Void radius [$h^{-1}\mathrm{Mpc}$] & No. of voids\\ \hline\hline
$ M_{{\rm halo}} < 10^{12.5}$ & 376233 & $5 \leq R_{{\rm void}} < 15$ & 109466 \\ 
$10^{12.5} \leq M_{{\rm halo}} < 10^{13.5}$ & 250362 & $15 \leq R_{{\rm void}} < 25$ & 3683\\ 
$M_{{\rm halo}} \geq 10^{13.5}$ & 25358 & $25 \leq R_{{\rm void}} < 35$ & 667\\ 
\hline
\end{tabular}
\end{center}
\end{table*}

\subsubsection{Doppler lensing algorithm}

\begin{figure}
    \hspace{-0.3cm}
    \centering
    \includegraphics[width=\columnwidth]{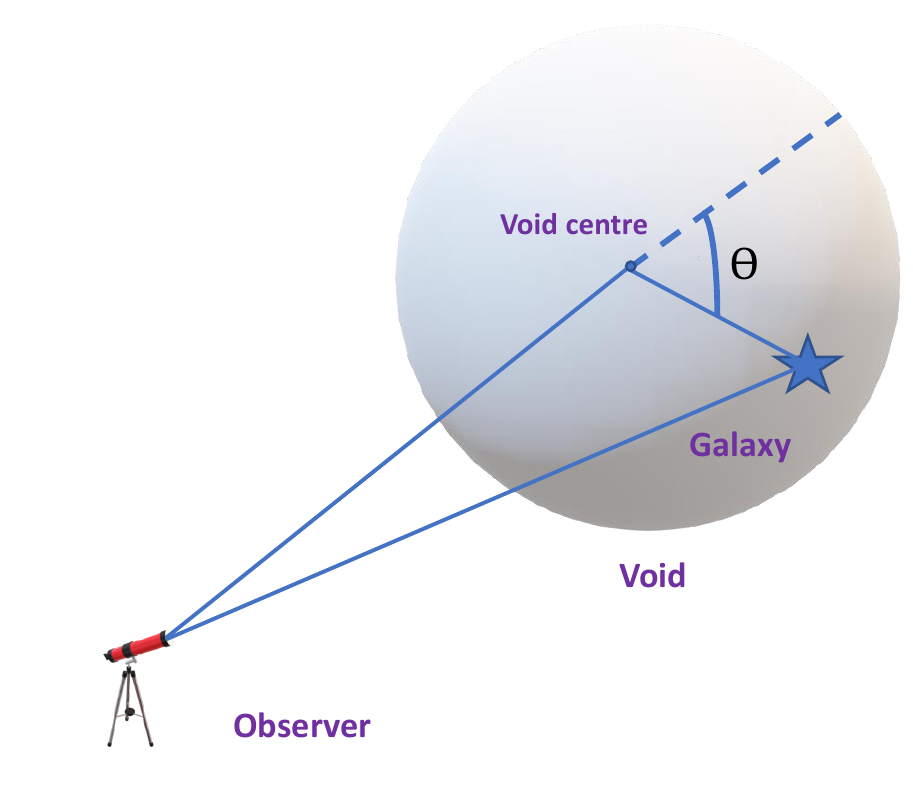}
    \caption{Schematic representation of the Doppler lensing observation for voids.
    }
    \label{figDLV}
\end{figure}

\begin{figure}
    \hspace{-0.3cm}
    \centering
    \includegraphics[width=\columnwidth]{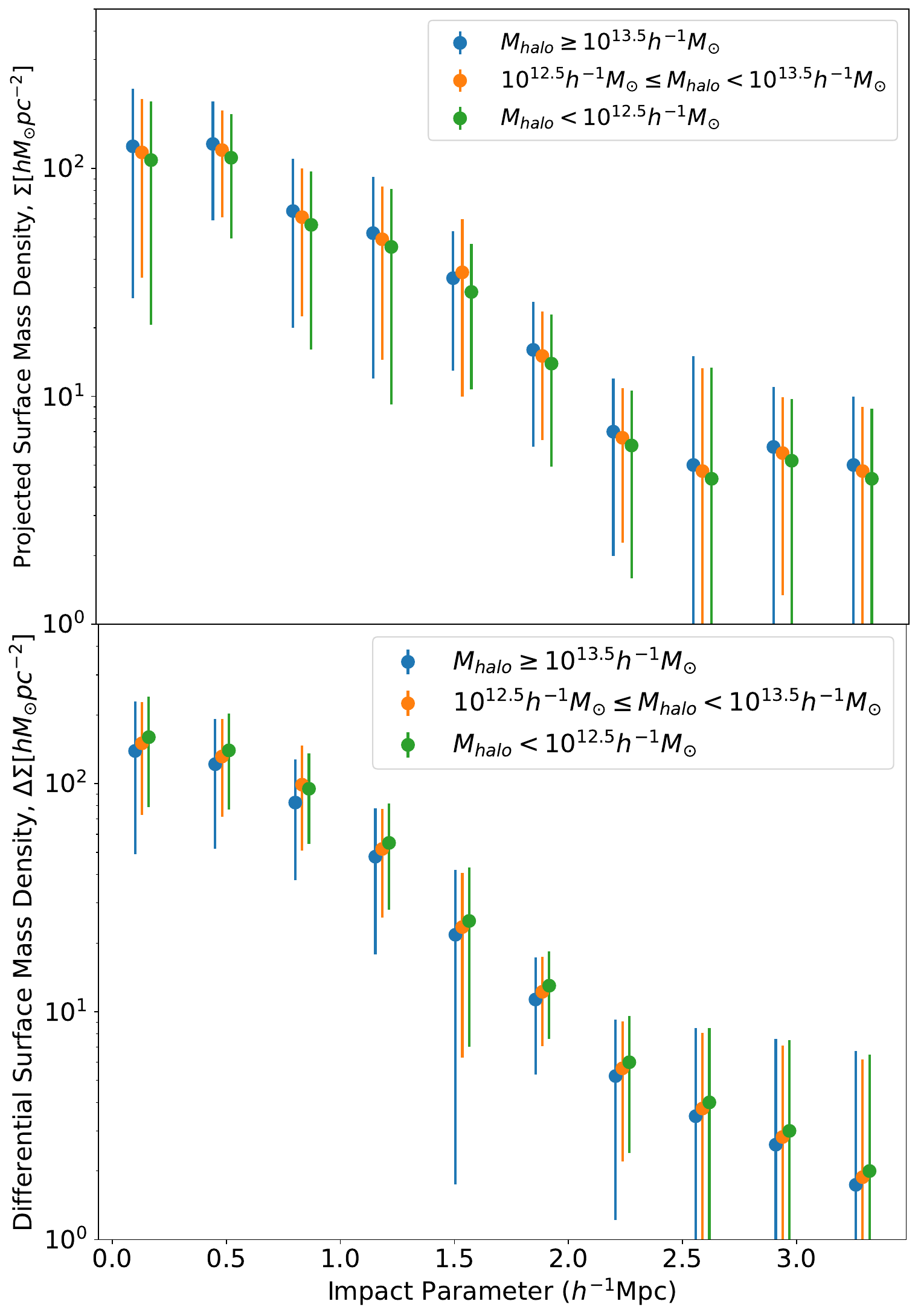}
    \caption{Variation of the weak-lensing statistical quantities as a function of the impact parameter (the distances from the centre of the halos to the centre of bundles of photons). The top panel shows the projected surface mass density (proportional to the weak-lensing magnification) as a function of the  impact parameter for clustering masses of the halos. The different colour dots indicate the different ranges of masses of the halos and the error bars represent the $1\sigma$ for each bin. The bottom panel shows the differential projected surface mass density (proportional to the weak-lensing shear) as a function of the impact parameter for clustering masses of the halos. We artificially shift the last two categories of halo masses bins to improve the visual clarity.
    }
    \label{fig3}
\end{figure}

\begin{figure}
    \hspace{-0.3cm}
    \centering
    \includegraphics[width=\columnwidth]{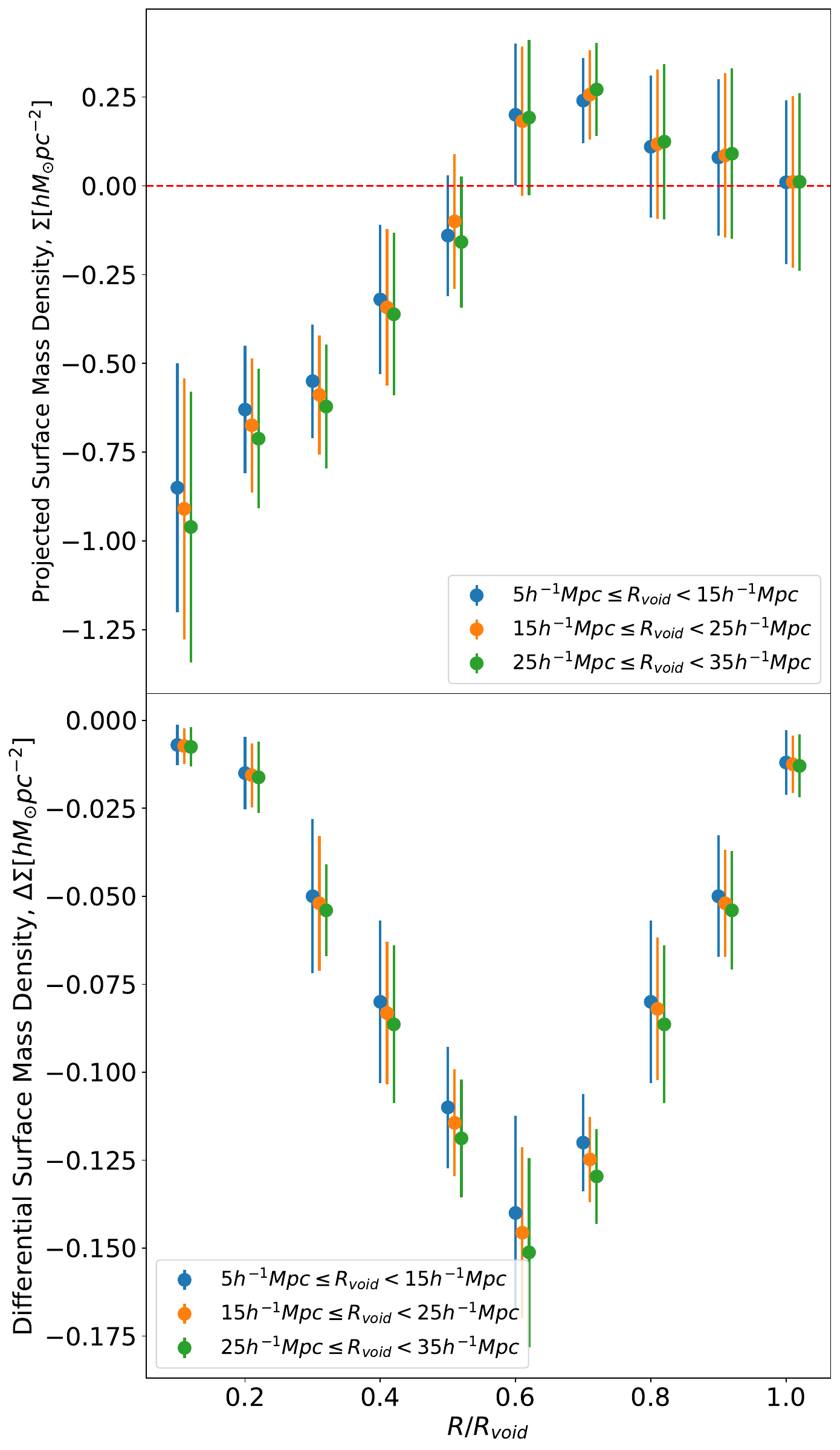}
    \caption{Variation of the weak-lensing statistical quantities as a function of the impact parameter (the distances from the centre of the voids to the centre of bundles of photons) normalise by the respective void radius. The top panel shows the projected surface mass density as a function of the  impact parameter for clustering radii of the voids. The different colour dots indicate the different ranges of radii of the voids and the error bars represent the $1\sigma$ for each bin. The bottom panel shows the differential projected surface mass density as a function of the impact parameter for clustering radii of the voids. We artificially shift the last two categories of void radii bins to improve the visual clarity.
    }
    \label{fig4}
\end{figure}

\begin{figure*}
    \centering
    \includegraphics[width=2\columnwidth]{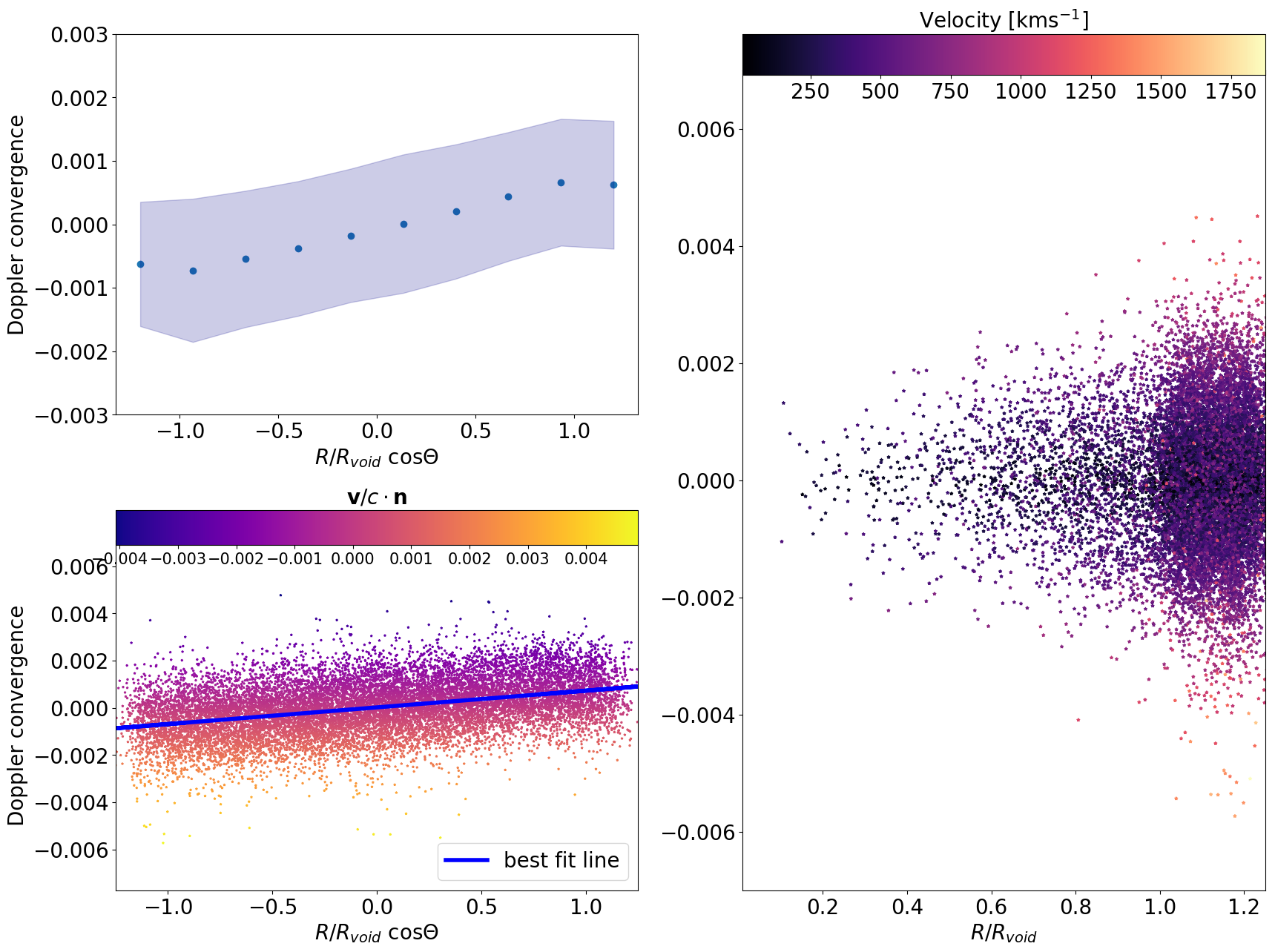}
    \caption{Doppler convergence by stacking 1433 voids having radii $20-25 ~h^{-1}\mathrm{Mpc}$ at $z = 0.22$. The top left figure shows the average Doppler convergence for a certain range of binned value of $R/R_{{\rm void}}$ $\cos \Theta$ with $1\sigma$ error bar (blue shaded region). The bottom left panel of the figure shows the Doppler convergence as a function of $R/R_{{\rm void}}$ $\cos \Theta$. The colour bar indicates the dot product between the velocity of the galaxies inside the voids and the directions of photon propagation. In the colour bar, the value of $\bm v/c\cdot\bm n > 0$ ($\bm v/c\cdot\bm n < 0$) indicates that the object travel towards (away) from us. The blue line is the best fit line of the data. The right panel of the figure shows the  Doppler convergence as a function of $R/R_{{\rm void}}$. The colour bar indicates the magnitude of the velocities of the galaxies inside the voids.}
    \label{fig5}
\end{figure*}

\begin{figure*}
    \centering
    \includegraphics[width=2\columnwidth]{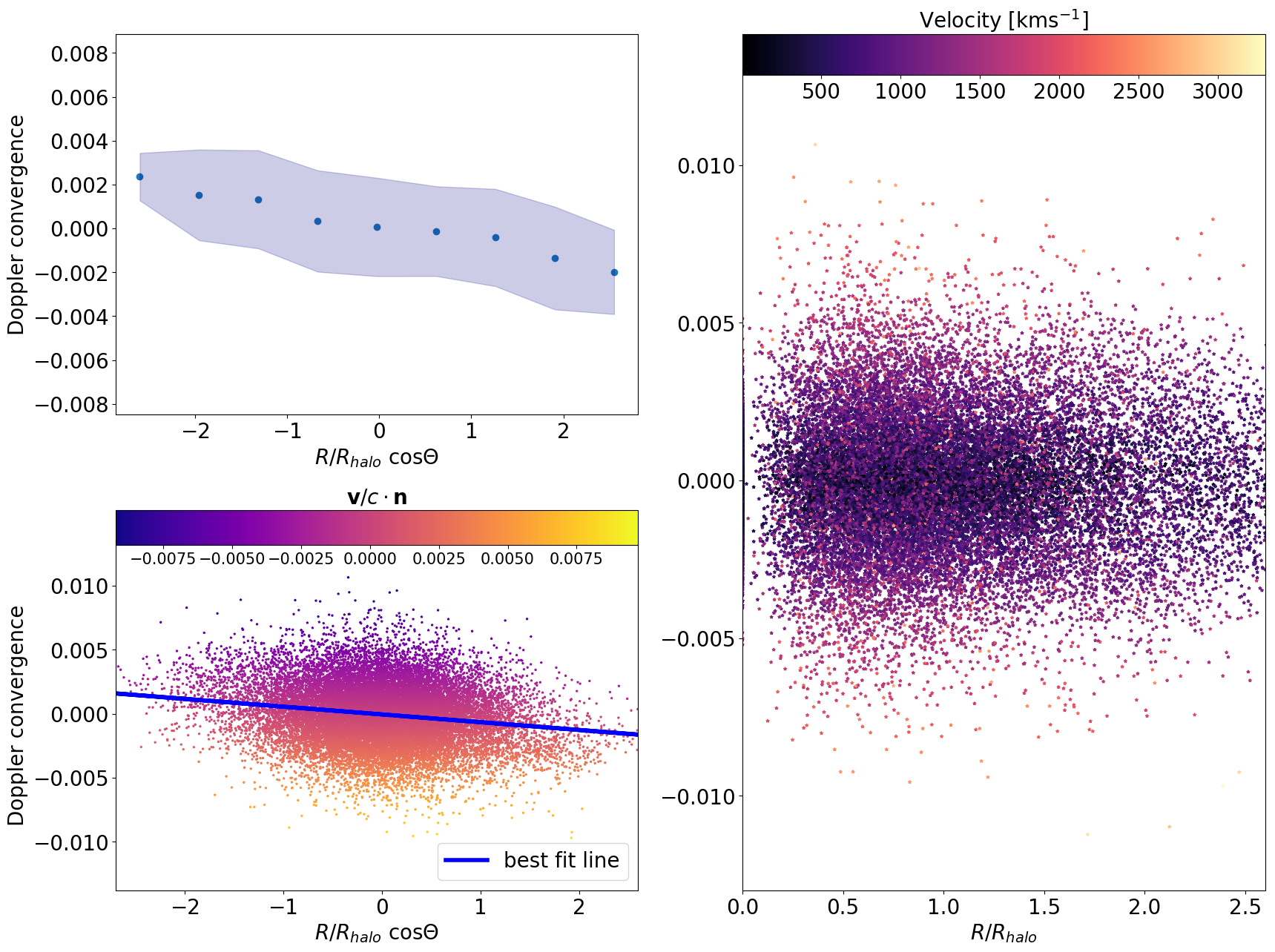}
    \caption{Doppler convergence by stacking halos at $z=0.22$. The top left figure shows the average Doppler convergence for a certain range of binned value of $R/R_{{\rm halo}}$ $\cos \Theta$ with $1\sigma$ error bar (blue shaded region). The bottom left panel of the figure shows the Doppler convergence as a function of $R/R_{{\rm halo}}$ $\cos \Theta$. The colour bar indicates the dot product between the velocity of the centre of the galaxies to the other galaxies and the directions of photon propagation. In the colour bar, the value of $\bm v/c\cdot\bm n > 0$ ($\bm v/c\cdot\bm n < 0$) indicates that the object travel towards (away) from us. The blue line is the best fit line of the data. The right panel of the figure shows the Doppler convergence as a function of $R/R_{{\rm halo}}$. The colour bar indicates the magnitude of the velocities of the galaxies. }
    \label{fig6}
\end{figure*}

To calculate the Doppler convergence, we solve Eqn. \ref{eqn:Dop} and analyse the statistical properties by stacking cosmic voids and halos. We follow the following steps to compute the Doppler convergence: 

\begin{enumerate}[i]
\item
 by changing the initial conditions we initially run 21 \texttt{gevolution} simulations (having identical cosmological parameter settings, but use a different random seeds) and then from the particle snapshots of \texttt{gevolution} simulations we find out the necessary information about the halos and voids at redshift $z= 0.22$ by using \texttt{ROCKSTAR} and \texttt{Pylians} (these are the same simulations as for the weak-lensing analysis); 
 \item the choice of 21 of snapshots was motivated by the need to simulate the volume of a slice between $z=0.2$ and $z= 0.25$. This is important for obtaining an unbias Doppler lensing signal for cosmic voids: the Doppler lensing is generated by peculiar velocities of observed galaxies, thus a minimal number of galaxies per an observed structure is required. For galaxy clusters this is not an issue (i.e. plenty of galaxies within large halos), however for voids this possesses challenges, especially for small voids where there can be no galaxies inside. For this reason we focus on  voids having radii $20-25$ Mpc$/h$ (as the number of halos inside the voids is very few for smaller voids so we consider a moderate range of void radius) at redshift $z=0.22$ and identify halos within these voids as well as halos beyond the radius of the associated voids (we find 1433 such voids);
 \item  then we compute the distances from the centre of the cluster/void to the centre of the observed halo;
 \item  we perform the dot product between the normalised velocities ($\bm{v}/c$) of the halos associated with clusters/voids and the direction vectors ($\bm{n}$) between the observer and the observed halos (cf. Fig. \ref{figDLV}); 
 \item then we multiply the rest of the parts of Eqn. \ref{eqn:Dop} with the result of the dot product and calculate the value of Doppler convergence;
 \item we stack Doppler convergence value for voids having radii $20-25$ Mpc$/h$, and using the same procedure we also computed the Doppler convergence by stacking halos;
 \item finally, when presenting results (cf. Figs. \ref{fig5} -- \ref{fig8}) we scale the radius by a factor of $\cos\Theta$.
Formally, speaking one would expect to see a plot of the Doppler convergence as a function of velocity  (cf. eq. \ref{eqn:Dop}), i.e. $ \kappa_v \sim \bm{v} \cdot \bm{n} \sim {v} \cos\Theta.$
 However, when presenting the results of the Doppler convergence, instead of ${v}$ we use $R$. The motivation is as follows: for a top-hat void the gravitational potential is quadratic, hence $ \phi \sim R^2~~ \Rightarrow~~ v \sim \nabla \phi \sim R,$
 i.e. velocity $v$ is linearly proportional to $R$. For more realistic voids, this is no longer true, but it provides a useful way of presenting results, i.e. as a function of distance from the center of the void (which makes it more comparable with results of the gravitational lensing). The other alternative would be to remove the factor of
 $\cos\Theta$ from the observed convergence by dividing the observed convergence the factor of   $\cos\Theta$ (i.e. $\kappa_{v}/ \cos\Theta$), this however would lead to problems with $\cos\Theta \approx 0$, and thus we do not re-scale the convergence by  $1/\cos\Theta$ by instead we re-scale the radius. Consequently, Figs. \ref{fig5} -- \ref{fig8} present the results as $\kappa_v \sim R\cos\Theta $.
 
\end{enumerate}

It is important to mention here that we develop the Doppler lensing algorithm that computes the Doppler convergence by considering fully three-dimensional (3D) signal stacking as a function of the radius of cosmic structures. Due to the consideration of 3D stacking, the smaller void consists of very few galaxies (or even completely empty) as compared to the larger void. Thus, even if the number of smaller voids
 are larger (cf. Fig. \ref{new_fig}), larger voids offer more halos per void.

\section{RESULTS AND ANALYSIS}\label{res}

To describe the results obtained from this work, at first we will discuss the weak-lensing with stacked clusters and voids, and then discuss the Doppler lensing with stacked clusters and voids. We will also discuss the comparison between Doppler lensing convergence and weak-lensing convergence.

\subsection{Weak-lensing with stacked clusters and voids}

How is the weak-lensing signal impacted by the different masses of the halos? To find out the answer we stacked the weak-lensing signal for clusters with various masses. We projected bundles of photons in the directions of halos with different masses and computed the weak-lensing convergence, shear and magnification by using our ray-tracing algorithm that we described in Section \ref{ray_alg}. For each mass range of the halo, we projected bundles of lights in the directions of galaxy cluster by slightly varying the projection angle and stacked 100 galaxy clusters to analyse the statistical property of the weak-lensing statistics. Then by using Eqn. \ref{eq:proj} we computed the projected surface mass density (which is proportional to the convergence) and differential surface mass density (which is proportional to the shear). Table \ref{tab:all_void_halo} lists the number of voids and halos from 21 realisations of \texttt{gevolution} simulations at redshift $z=0.22$.

In Fig. \ref{fig3}, the top panel shows the variation of the convergence as a function of impact parameter (the distance from the centre of the halos to the centre of the bundles of photons). The numerical result of the weak-lensing magnification is higher near the centre of the halos while it gradually decreases with the increase of impact parameter. 
As expected the weak-lensing magnification is slightly higher when the bundle of light passes near the massive halos than the less massive halos.
The variation of the weak-lensing tangential shear as a function of impact parameter is depicted in the bottom panel of Fig. \ref{fig3}. Similar to weak-lensing magnification we observed that the value of the tangential shear is maximum near the centre of the halos and gradually decreases when the distance from the centre of the halo to the centre of the bundle increases. We also observed that the characteristic plot of the tangential shear is opposite to that of the weak-lensing magnification, with smaller halos having a slightly higher amplitude. As we have considered the absolute magnitude value of the shear so the characteristics, what we observed for shear, are exactly opposite that we observed for weak-lensing magnification. Our result is also consistent with the result of \cite{Schmidt2012} (cf. their Fig. 4) where they have considered the effects of galaxy sizes and magnitudes on the measurement of weak-lensing magnification.

\begin{table}
\begin{center}
\begin{tabular}{@{}llcccc@{}}
\toprule
   & Void radius [$h^{-1}\mathrm{Mpc}$]                        & \multicolumn{1}{l}{No. of voids} & \multicolumn{1}{l}{No. of halos} & Halos per void \\ \midrule
& $20-25$                 & 1433                               & 9895                                  & 7            \\
& $25-30$       & 547                               & 7581                                   & 14               \\
& $30-35$  & 227                                & 6051                                  & 27               \\
\bottomrule
\end{tabular}
\caption{Doppler lensing with stacked voids: number of voids in each radial bin from 21 realisations of \texttt{gevolution} simulations at redshift $z=0.22$. The information of voids and halos are extracted from the publicly available codes \texttt{Pylians} and \texttt{ROCKSTAR}, respectively. The third column shows the number of halos inside the corresponding number of voids and the fourth column shows the average number of halos per void.}
\label{table:dop_tab}
\end{center}
\end{table}

The projected surface mass density and differential surface mass density for voids with different radii are depicted in Fig. \ref{fig4}. The value of the magnification and shear is negative inside voids due to the low-density distribution. 
The key difference between the gravitational lensing with stack clusters and the gravitational lensing with stack voids is the relative difference between the convergence signal and shear signal. For clusters, the projected mass density ($\Sigma$ which is proportional to lensing convergence) and differential surface mass density ($\Delta \Sigma$ which is proportional to shear) are of the order of $100 M_\odot pc^{-2}$. For voids the convergence is small ($\Sigma \sim 1 M_\odot pc^{-2}$) is even smaller but the shear ($\Delta \Sigma \sim 0.1 M_\odot pc^{-2}$). Still, as seen from Figs.  \ref{fig3} and  \ref{fig4} 100 objects will suffice to detect a weak-lensing signal. This should be contrasted with the results of the next section where the detection of the Doppler lensing at this redshift requires a factor of 10 large number of objects ($\sim 1000$). 

Finally, it is also apparent from Fig. \ref{fig4} that a void finding algorithm overestimates the apparent size of a void (i.e. a void is approximated with a sphere and the radius of a sphere is treated as $R_{void}$). The apparent edge of the void (i.e. as observed on the sky) coincides with the maximum of the convergence and minimum of the shear, which as seen in Fig. \ref{fig4} is approximately at $R/R_{void} \approx 0.6-0.7$. 
Thus, the actual size of the void (as inferred from the shape of the lensing signal) does not coincide with the value of the `radius' produced by the implemented void-finder.
While more robust void finding algorithms \citep{Nadathur2015a, Nadathur2015b, Sutter2015} does not suffer such a problem, it needs to be noted that the nominal value of the radius of the void does not affect the discussed result. The ray-tracing algorithm is not sensitive to such a scaling (it is only affected by the matter distribution along the line of sight, not by any artificial definition of the radius of a void).


\subsection{Doppler lensing with stacked voids and halos}\label{res:Dop}

Here we will discuss some results of Doppler lensing and how it differs from the standard weak gravitational lensing effect. Figure \ref{fig5} shows the variation of Doppler convergence by stacking 1433 voids having radii $20-25 ~h^{-1}\mathrm{Mpc}$. The top left of Fig. \ref{fig5} shows the mean Doppler convergence (blue dots) for certain ranges of distance bins and the blue shaded region shows the $1\sigma$ variation of data. The bottom left figure shows the Doppler convergence as a function of $R/R_{{\rm void}}$ $\cos \Theta$, the blue line is the best fit line of the Doppler convergence data, and the colour bar indicates the dot product of the velocities of the galaxies (normalised by the speed of light) and the propagation directions of the photons. The value $\bm v/c \cdot\bm n > 0$ ($\bm v/c \cdot\bm n < 0$) in the colour bar indicates that the object travel towards (away) from us. The `+ve' (`-ve') sign of the value of Doppler convergence implies that at their observed redshift they are smaller and dimmer (larger and brighter) than typical objects \citep{Bolejko2013, Bacon2014}. 
We emphasise that $\bm v$ is a $3D$ vector and hence the stacking of the Doppler lensing signal is a full 3D stacking.

From Fig. \ref{fig5} it is apparent that the value of Doppler convergence due to underlying matter distributions in the void increases as the distance between the center of the void and the galaxies within the void increases, similar to standard weak-lensing statistical quantities. It is also clear that the observer could measure the magnified image of the source at the edge of the void for Doppler lensing analysis in comparison to the result of standard weak-lensing statistics. The reason for the higher Doppler convergence value at the void's edge is that the velocity distribution (since the Doppler convergence is proportional to the velocity of the source galaxies) is higher there than in the centre and the right side of Fig. \ref{fig5} indirectly supports the validation of the statement where the variation of Doppler convergence as a function of $R/R_{{\rm void}}$ is depicted. Another reason for getting higher Doppler convergence value at the void's edge is higher particle density distribution there than in the centre, and the right side of Fig. \ref{fig5} implicitly supports the validation of the argument (as most of the galaxies inside the voids are situated near the edge of the voids than the centre) and the colour bar indicates the velocities of the galaxies inside the voids.
A similar figure but for stacking halo analysis is depicted in Fig. \ref{fig6}. It is found that the Doppler convergence properties by stacking halos are very close to that of standard weak gravitational lensing. Since the halos have a smaller radius as compare to voids so we consider the effect of the galaxies more than twice the radius of the halos from its centre to compute the Doppler convergence.

\begin{figure}
  \centering
  \noindent
  \resizebox{\columnwidth}{!}{
  \includegraphics[width=\columnwidth]{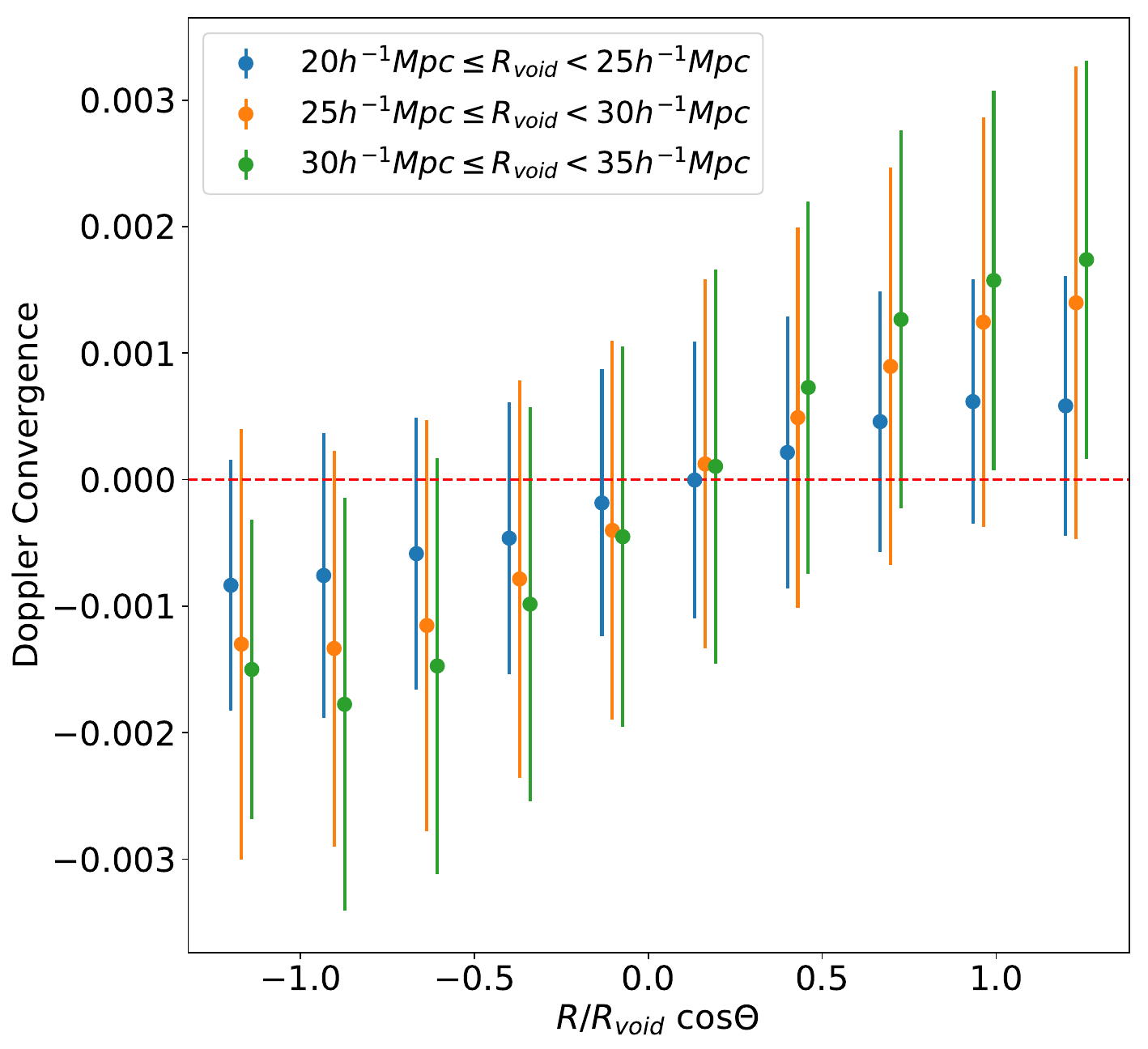}
  }
  \caption{Variation of the Doppler convergence as a function of the impact parameter for clustering radii of voids. The positive (negative) value of Doppler convergence indicates the object moving towards us (away from us). The different colour dots indicate the different ranges of radii of the voids and the error bars represent the statistical error for each bin. We artificially shift the last two categories of void radii bins to improve the visual clarity.}
  \label{fig7}
\end{figure}

Table \ref{table:dop_tab} lists the number of voids and the associated halos from 21 realisations of \texttt{gevolution} simulations for different ranges of radii of voids at redshift $z=0.22$. Figure \ref{fig7} shows the variation of Doppler convergence as a function of $R/R_{{\rm void}}$ $\cos \Theta$ by stacking voids having different ranges of void radii (void radii: see Table \ref{table:dop_tab}). To generate Fig. \ref{fig7}, we have considered all the voids for the respective ranges of void radii and obtained the Doppler lensing signal by stacking voids. It is found that Doppler convergence for different ranges of radii of voids is higher at the void edges and the observer could also measure the magnified image of the source for voids with larger radii than the smaller ones. Because, the number of galaxies in larger voids is greater than in smaller voids, and the velocity distribution and density distribution of galaxies are higher at the void's edge.  

\begin{figure}
  \centering
  \noindent
  \resizebox{\columnwidth}{!}{
  \includegraphics[width=\columnwidth]{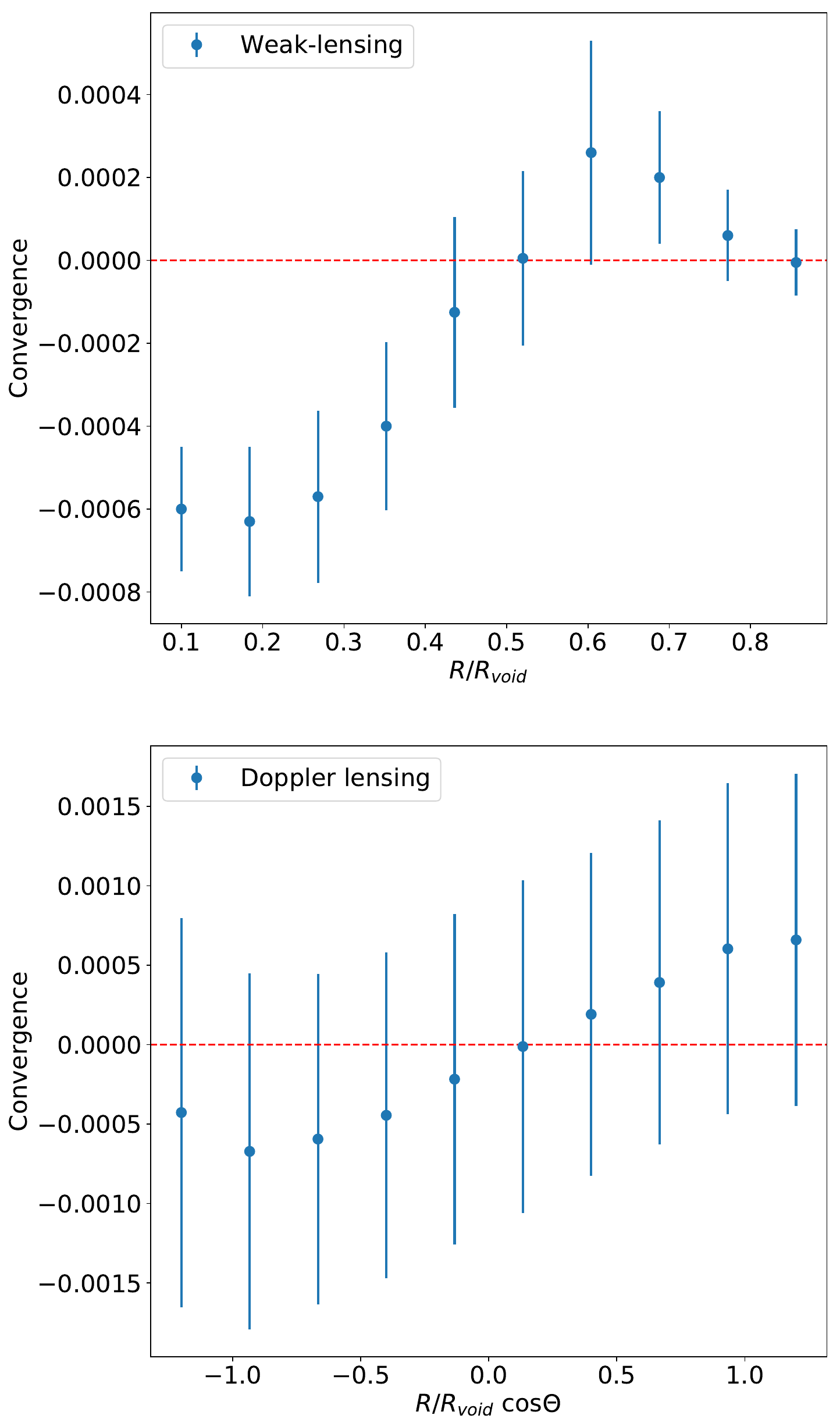}
  }
  \caption{Variation of the weak-lensing convergence and Doppler convergence as a function of the impact parameter by stacking voids having radii $20 h^{-1}\mathrm{Mpc} \leq R_{{\rm void}} < 25 h^{-1}\mathrm{Mpc}$.}
  \label{fig8}
\end{figure}

\begin{figure}
  \centering
  \noindent
  \resizebox{\columnwidth}{!}{
  \includegraphics[width=\columnwidth]{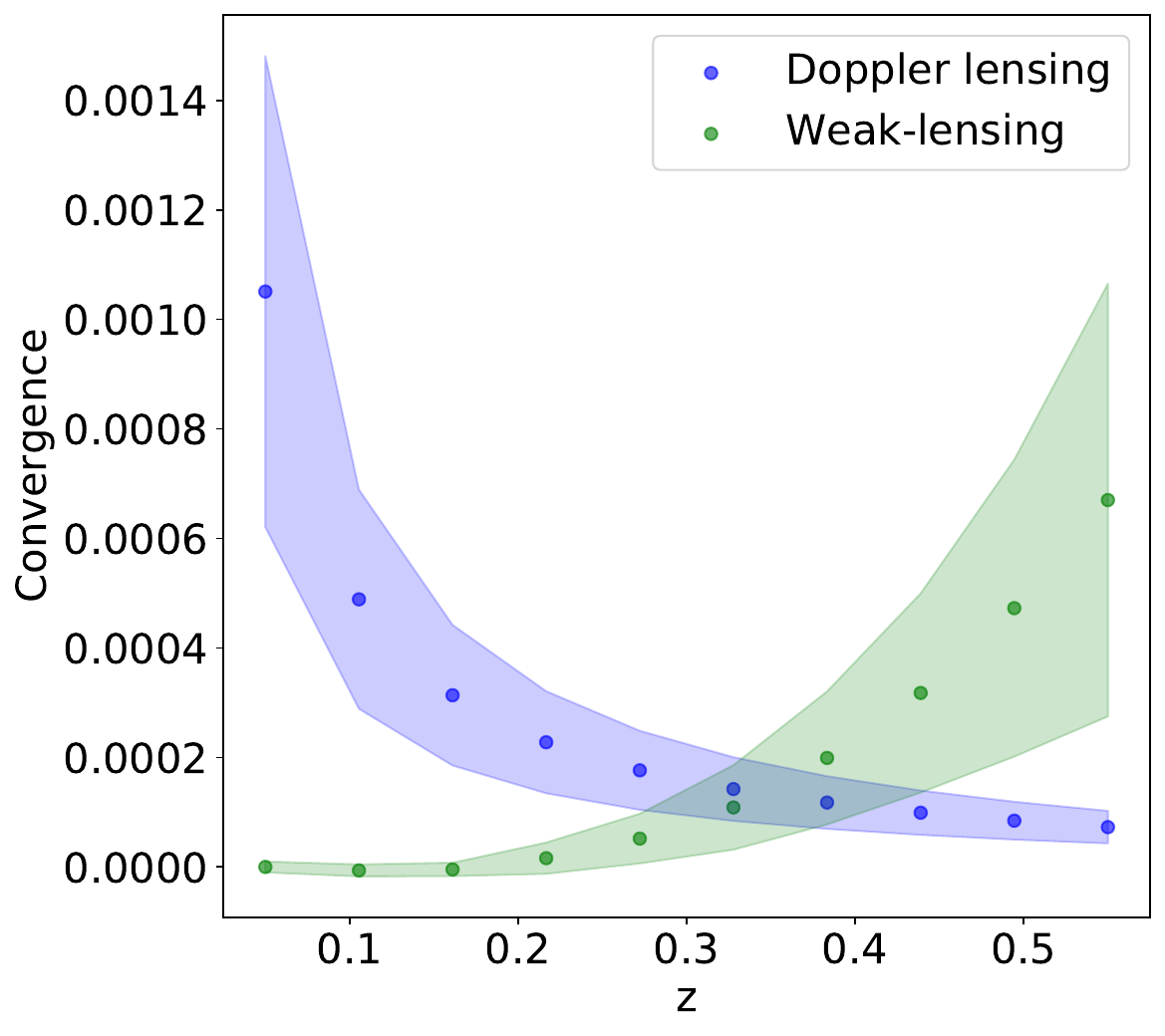}
  }
  \caption{A comparison between Doppler lensing convergence and the weak-lensing convergence (both for stacking halo analysis) as a function of redshift. For weak-lensing analysis, we have considered 100 halos having mass $M_{halo} > 10^{13.5}~h^{-1}M_\odot$. For Doppler lensing analysis, we have considered all of the halos having similar mass range but radius $R_{halo} > 1~ h^{-1}\mathrm{Mpc}$. The blue dots represent the mean Doppler convergence and the green dots represent the mean weak-lensing convergence. The shaded regions indicate 1$\sigma$ variation of the statistical error for each bin.}
  \label{fig9}
\end{figure}

Finally, Fig. \ref{fig8} shows the variation of the weak-lensing convergence and Doppler convergence as a function of impact parameter by stacking voids. To compare the weak-lensing and Doppler lensing statistics as a function of redshift, we have projected the bundles of photons in the direction of the centres of the halos and stacked them to analyse the statistical properties of both lensings. A comparison for Doppler lensing convergence and weak-lensing convergence as a function of redshift by stacking halos is depicted in Fig. \ref{fig9}. We have stacked weak-lensing and Doppler lensing signals by taking halos for different snapshots of \texttt{gevolution} simulations at different redshifts. The shaded regions (blue and green) indicate 1$\sigma$ variation for each redshift bin of the data. At lower redshift, the convergence value due to Doppler lensing dominates over the standard weak gravitational lensing, whereas at higher redshift, the convergence value due to Doppler lensing drops (since the factor in brackets in Eqn. \ref{eqn:Dop} decreases in amplitude) while the convergence value due to weak-lensing increases. Our result is also compatible with the findings of \cite{Bolejko2013} and \cite{Bacon2014}.

\section{Conclusions and Outlook}\label{conclusion}

This paper has numerically investigated signatures of weak gravitational lensing and Doppler lensing within relativistic $N$-body simulations.
We used the code \texttt{gevolution} to produce snapshots both in terms of particle distribution as well as relativistic gravitational potentials.
The novelty of this paper is that we implemented the RBM to integrate the null geodesics backward from the observer to the sources to obtain the path of the photon.
Although the RBM has been implemented in the past within $N$-body simulations, it is the first time it has been implemented in relativistic simulations. In addition, we also implemented the Doppler lensing method into our analysis. 

In our analysis, we considered various categories of halos and voids
and utilized information from statistical quantities such as convergence, shear, and magnification. For halos, we considered three categories of halo masses and stacked them to extract the cosmological information from weak-lensing and Doppler lensing analysis.
For voids, we considered three categories of void radii and stacked them to extract the cosmological information from weak-lensing and Doppler lensing analysis. 

The results for halos for the gravitational lensing were as expected: the weak-lensing magnification and shear are higher near the centre of the halos while it gradually decreases with the increase of impact parameter. We also observed a slight variation of the measurement of weak-lensing magnification and shear for different categories of halo masses. 
For voids, we observed the amplitude of the convergence to be the largest at the centre while the shear was the highest near the edge of voids.
Qualitatively, such an outcome is not novel as it was predicted by other methods, but the importance of the results presented in this paper is that the analysis was done based on relativistic simulations rather than relied on Newtonian $N$-body simulations or perturbative methods.

The most important result of this paper was the analysis of Doppler lensing.
We examined the Doppler lensing statistics for stacked halos and voids. 
For halos, we showed that the characteristic results of Doppler convergence by stacking halos are very close to that of standard weak gravitational lensing. Since the Doppler lensing is sourced by peculiar velocities, the Doppler lensing signal averages out near the center of halo. It is only at the outskirts of these halos when a slight trend can be observed, cf. Fig. \ref{fig5}. We speculate that this effect could in principle affect type Ia Supernovae and could potentially lead to a weak correlation between the cosmological environment of a host galaxy and the peak magnitude of the observed supernova, in a similar way as one observes a weak correlation between the peak magnitude and the mass of a host galaxy  \citep{Sullivan2010}. Similarly, the investigation of the Doppler lensing signal produced by cosmic voids rendered interesting results. The most important one was related to difficulties in extracting the Doppler lensing signal produced by cosmic voids.

The distinct difference between gravitational lensing and Doppler lensing is the origin of the signal. For gravitational lensing, an observed galaxy is far beyond the lens that produces the gravitational lensing signal. Thus, for gravitational lensing, the precise location of an observed galaxy is less relevant and hence one can always find a suitable number of objects behind the lens to measure and analyse the gravitational lensing signal. For the Doppler lensing signal, however, the source is at the "lens". While for clusters there are ample galaxies, for voids this causes potential problems with detectability. 
Since there is a very little number of galaxies inside voids, we had to consider voids with a radius between  20 and 25 $h^{-1}$ Mpc. To detect a sufficient number of such voids one requires large volumes. However, larger volumes imply larger redshifts, which in turn leads to a lower amplitude of the Doppler lensing. As the amplitude of the Doppler lensing decreases with redshift, rather than increases as it is the case of the gravitational lensing, we had to settle with the redshift of around $z=0.2$ (to be precise $z =0.22$).
In our studies, we found that we required more than 1000 voids to obtained a stacked signal that could in principle be measured. 
For this we had to simulate the volume of approximately ($900~h^{-1}$ Mpc)$^3$, and within this volume we identified the total number of 1433 voids with a radius between 20 and 25 $h^{-1}$ Mpc.
The considered volume corresponds to a volume between redshift slice of $z=0.2$ and $z=0.25$.

By stacking 1433 voids, we showed how galaxies inside these voids are distributed and concluded that the majority of galaxies within the voids are closer to the edges than the centres. We demonstrated that Doppler lensing predicts that the observer could see the magnified picture of the source at the void's edge, despite the fact that the standard weak-lensing fails to predict. In this work, we have also demonstrated the comparison between standard weak gravitational lensing and Doppler lensing. For structures (i.e. "lenses") at lower redshifts, the convergence is dominated by the Doppler lensing, whereas if the lens is at high redshift the convergence is dominated by the gravitational lensing.

The results of this paper also indicate that while the optimal redshift for measuring the Doppler lensing on voids is approximately $z=0.2$,
the most optimal survey strategy (targeting the same structures and taking use of both Doppler lensing and gravitational lensing) should focus on redshift around $z_{lens} \approx 0.3-0.4$.
At this redshift range there is a sufficient number of voids/halos to take advance to this new strategy based on both the weak gravitational lensing and Doppler lensing. The strategy would aim at targeting galaxies around the halos/voids ($z_{source} \approx 0.3-0.4$) with the aim to extract the Doppler lensing signal and background galaxies ($z_{source} \approx 0.6-0.8$) with the aim to extract the gravitational lensing signal. With the gravitational lensing the signal is mostly encoded in the amplitude and to some extent in the shape, cf. Figs. \ref{fig3} and \ref{fig4}; whereas with the Doppler lensing signal the signal can be extracted from the slope, cf. Figs. \ref{fig5} and \ref{fig6}.

Cosmic voids occupy approximately 70\% of our universe with very little content of baryonic matter. The scarcity of baryonic matter inside voids (which means very little contamination from complex baryonic astrophysics) makes the voids pristine environments, which are ideal to test properties of dark matter and the nature of gravity. Over the recent years, there has been a growing interest in investigation properties of cosmic voids  \citep{Park2007, Bos2012, Pan2012, Krause2012, Sutter2012, Bolejko2013, Ceccarelli2013, Hamaus2014, Nadathur2014, Hamaus2015, Nadathur2016, Sanchez2016, Mao2017, Verza2019, Panchal2020, Li2020, Raghunathan2020}. 
The results of this paper should be treated as a proof-of-concept that investigated and showed the possibility of using the Doppler lensing in conjunction with the gravitational lensing to map the matter distribution inside cosmic voids.
The results presented here are relevant to ongoing and future low-redshift spectroscopic surveys such as, for example, the DESI Bright Galaxy survey \citep{Ruiz-Macias2020, Zarrouk2021}. Even though DESI is not a lensing survey, it will provide spectroscopic data that will be used for Doppler lensing, which will be utilised in conjunction with gravitational lensing, and other methods in order to better map the underlying matter distribution within cosmic voids.
In our future works, we will combine the standard weak gravitational lensing and Doppler lensing signals to extract the DM profiles, and also consider the higher-order statistics i.e. angular power spectrum, bi-spectrum, etc. due to underlying matter distributions in and around the cosmic voids/halos at higher redshift. In addition, the covariance effects on the joint measurement of the Doppler lensing statistics using halos and voids, and the constraints of cosmological parameters due to these lensing effects would be interesting and we left it for our future works.  

\section*{Acknowledgements}

The authors would like to thank Julian Adamak for his discussions during the early stages of this work, and Florian List for his reading and comments on this paper. We further acknowledge the use of Artemis at The University of Sydney for providing HPC resources that have contributed to the research results reported within this paper. The authors thank the \texttt{gevolution} team for making the code publicly available, and the anonymous referee for providing useful remarks that contributed to the final form of this paper.
MRH is supported by the Australian Government Research Training Program (RTP) Scholarships.
MRH would like to thank Lawrence Dam and William H. Oliver for their discussions during this work.
KB acknowledges support from the Australian Research Council through the Future Fellowship FT140101270. This research work made use of the free Python packages \texttt{numpy} \citep{Harris2020}, \texttt{matplotlib} \citep{Hunter2007}, \texttt{h5py}\footnote{\orange{https://www.h5py.org/}},  \texttt{eqtools}\footnote{\orange{https://eqtools.readthedocs.io/en/latest/}}, and \texttt{mpi4py} \citep{Dalcin2005}.

\section*{DATA AVAILABILITY}
The data generated as part of this project may be shared with the corresponding author upon reasonable request. 









\bsp	
\label{lastpage}
\end{document}